# Effects of Composition on Mechanical and Antibacterial Properties of Hydroxyapatite/Gray Titania Coating Fabricated by Suspension Plasma Spraying

Md Mirazul Mahmud Abir[a], Yuichi OTSUKA[b,], Kiyoshi OHNUMA[c], Yukio MIYASHITA[d]

[a]*Graduate School of Information and Control Engineering, Nagaoka University of Technology, 1603-1 Kamitomioka, Nagaoka-shi, Niigata 940-2188, Japan.*
[b]*Department of System Safety, Nagaoka University of Technology, 1603-1 Kamitomioka, Nagaoka-shi, Niigata 940-2188, Japan.*
[c]*Department of Biongineering, Department of Science of Technology Innovation, Nagaoka University of Technology, 1603-1 Kamitomioka, Nagaoka-shi, Niigata 940-2188, Japan.*
[d]*Department of Mechanical Engineering, Nagaoka University of Technology, 1603-1 Kamitomioka, Nagaoka-shi, Niigata 940-2188, Japan.*

Abstract

This study aims at revealing the effect of composition on mechanical and antibacterial properties of hydroxyapatite/gray Titania coating for biomedical applications. HAp is a bioceramic material used as a plasma-sprayed coating to promote osseointegra-tion of femoral stems. Biomaterial coatings are fabricated mostly using atmospheric plasma spray (APS). However, the conventional plasma spray process requires hydrox-yapatite powder with good flow, and to prepare such free-flowing powder, agglomer-ation, such as spray drying, fusing, and crushing is used. Therefore, it is impossible to spray nano-powder using conventional methods. Here, we designed a suspension-feeding system to feed nanoparticles using a liquid carrier. Suspension plasma spray (SPS) successfully deposited homogeneous HAp/Gray Titania coating with less poros-ity on the surface of titanium substrates. The microstructure of coatings with differ-ent compositions was then characterized using scanning electron microscopy, X-ray diffraction, and Raman spectroscopy to identify the crystal structure. All results con-sistently demonstrated that SPS could transform $Ti_2O_3$ into $TiO_2$ with mixed Magneli phases, such as $Ti_4O_7$ and $Ti_3O_5$, which usually show photocatalytic activity. Inter-

Corresponding author: otsuka@vos.nagaokaut.ac.jp

facial strength, hardness, young modulus, and fracture toughness were also improved with a high concentration of $TiO_2$. Antibacterial test for *E.Coli* under LED light re-vealed that SPSed HAp/Gray Titania coating could significantly enhance antibacterial properties. The possible underlying mechanism of enhanced antibacterial properties can be attributed to increased Magneli phases and better bacterial adhesion caused by hydrophilic properties due to submicron size particles. SPS can fabricate visible light-responsive antibacterial coating, which can be used for medical devices.

1. Introduction

Aging people are suffering from locomotive diseases, such as osteoarthritis or as-sociated symptoms in the bone tissue and joints [1]. Metallic implants to replace the deteriorated human bone have been widely used [2]. Commercially available pure ti-tanium (cp-Ti ASTM F16) and titanium alloy Ti-6Al-4V ELI (extra-low interstitial) are the typical titanium alloys for biomedical applications. However, such the metal-lic biomaterials do not have any bioactivity to form a bond with surrounding tissues, which causes lower adhesive strength [1]. To overcome insufficient bonding, bioactive, as well as durable coating should be deposited onto the surface of metal implants [3].

Bioactive ceramics, which can promote osseointegration and osteoconduction, can form strong bonds and lead to enhanced durability of the implant when it is subjected to mechanical loading [3]. Hydroxyapatite [$Ca_{10}(PO_4)_6(OH)_2$, HAp], the most promi-nent bioactive ceramic which has a similar chemical composition to the human bone, has been widely used as an artificial bone or for coating metal implants to promote osteoconductivity [4]. However, HAp is a brittle ceramic material and then cannot be applied to a highly stressed area [5]. Development of HAp-based composite ma-terials with Ti$O_2$, Yttria Stabilized Zirconia, Alumina, Nanodiamond, Magnesium, or natural fiber, has been actively investigated to enhance mechanical properties of HAp [6, 7, 8, 9]. Mechanical properties of plasma-sprayed HAp coating was also reported and it is highly sensitive to crystallinity and the phase of HAp because its crystal struc-ture changes to another phases of calcium phosphate during heating process [10]. Fatigue crack propagation of plasma-sprayed HAp coating is sensitive to cyclic loading and its microstructure should be considered to enhance its durability [3, 11, 12, 13].

Another considerable function of the bioactive coating is antibacterial properties [14]. Infection of the implant due to bacterial adhesion is one of the main causes of implant failure [15]. Complications, such as tooth decay after long-term use or the need for joint replacement surgery, sometimes occur due to bacterial infection [14]. Several antibacerial coating methods have been proposed [16, 17], such the addition of the Ag ion [18], antibiotic peptide [19], and organic compounds [20]. A concept of multifunctional coating has recently been proposed [14], and coating with antibacterial properties, as well as high biocompatibility, is indispensable [21]. One possible multifunctional plasma-sprayed coating is a composite of HAp with a photocatalyst, such as Ti$O_2$, which provides a controllable release of radicals upon UV irradiation [22, 23, 24]. Unfortunately, UV irradiation is harmful to organs and visible light is more preferable. A composite plasma-sprayed coating of HAp with a visible light-responsive photocatalyst (Ti$O_2$-containing Magneli phases called "gray titania") has been successfully produced [25]. However, the composite HAp/Gray Titania coating has a relatively low bacterial adhesion and antibacterial properties, so the HAp complex [26, 27] was used to enhance cell adhesion, which demonstrated enhanced antibacterial property upon visible light irradiation [25, 21, 28]. The deficiency can be attributed to heating process during plasma spraying, which leads to low crystallinity of HAp thermal decomposition into more soluble phases ( -TCP) [29, 30, 31]. Plasma spraying should be further considered to improve antibacterial properties of plasma-sprayed HAp composite coating.

Thermal spray technology, particularly atmospheric plasma spray (APS), is the accepted method by Food and Drug Administration (FDA, USA) to fabricate HAp coat-ing for biomedical applications [32, 16, 17]. To suppress thermal decomposition, sev-eral spraying technologies have been developed, like vacuum plasma spray [33], cold spray [34], powder jet deposition [35], or aerosol deposition technology [36]. Vacuum thermal spray technology can reduce oxidation but the productivity is limited [33]. Cold spraying was also proposed to prevent particle melting during spraying. The method was considered not applicable for ceramics [34]. Recently, jet deposition using fine particles in a vacuum was successfully applied to fabricate thin ceramics coatings [35, 36, 37]. However, the method is challenging when applying a thick coating in a short period. Suspension plasma spray (SPS) is an emerging coating technique that uses feedstock in the form suspension instead of dry powder, and therefore, it is ca-pable of depositing nano-sized particles [38]. Poor flow of the ultrafine particle can be overcome by using a suspension flow system which can take particles to the center of the plasma jet [39, 40, 41]. Lower thermal decomposition is also expected due to the heat consumption effect by the solvent, which can lead to higher crystallinity of the HAp-based biomaterial coating [42]. However, the phase transformation of HAp and titania coating via SPS has not been studied and the effect of the composition on mechanical and antibacterial properties of SPS-HAp composite coating is not known.

This study aims at revealing the effect of composition on mechanical and antibacterial properties of HAp/Gray Titania coating for biomedical applications. Coating characterization was performed to investigate the microstructure and crystal structure. Scanning electron microscopy (SEM) was used to check the morphology of the coated specimen. Phase identification was performed using Horizontal X-ray diffraction (HXRD) and Raman spectroscopy. Interfacial strength was determined by applying a tensile test using acoustic emission (AE) measurement system. Fracture toughness, hardness and young modulus of each type of coating were also evaluated to observe the effect of composition on mechanical properties. Finally, antibacterial proper-ties of HAp/Gray Titania coating was evaluated by determining the number of colony forming units (CFU).

2. Experimental procedure

2.1. Preparation and characterization of powders

HAp powder (HAp-100 Taihei Chemical Industrial Co. Ltd., Japan) and titanium (III) Oxide ($Ti_2O_3$, Wako Chemical Industry Co. Ltd., Japan) were used as feedstock powders to deposit suspension plasma-sprayed coatings. Both types of powders were ball milled for 72 h at 75 rpm using ceramic balls of 20-mm diameter and were subse-quently sieved using # 45 m. The sieved particles were dispersed ultrasonically in the solution using a homogenizer for 20 min at a concentration of 30 wt.%. Distributions of particle size were measured using a laser diffraction particle size analyzer (LS-1300, Beckman Coulter, Inc., USA.). Morphologies of the powders were observed using a digital microscope (VHX-1000, Keyence CO. Ltd., Japan).

*2.2. Preparation of the suspension*

A mixture of water and ethanol at 1:1 ratio was selected, as shown in Table 1. Before adding the powder into the solvent it was measured to keep a concentration of 25 wt. % The suspension was stirred for 3 h at 350 rpm using a mixer to disperse the particle in suspension. The suspension was then filtered through a sieve of 45 m mesh to further segregate the larger particles to avoid clogging during spraying. To improve the durability of the suspension, dispersing agent sodium hexametaphosphate at 0.8 wt. % was mixed with the suspension just before coating fabrication.

*2.3. Material preparation and surface modification*

Commercially pure titanium (cp-Ti) at a rectangular shape and Ti-6Al-4V ELI at a cylindrical bar shape were used for coating using SPS. The rectangular plates were used to determine the optimum condition for the suspension flow system, coating characterization, and the antibacterial test. The cylindrical specimens were used to determine the interfacial strength of the coating using tensile test under monotonic loading. The rectangular plates with at a size of 10 x 50x t3 $mm^3$ were cut using a saw and, subsequently exposed to a milling machine and fine cutter to fit the specimen properly into the jig. The round bar specimen at a length of 60 mm and a diameter of 15 mm was prepared using a lathe machine and fine cutter. After completing all these processes, the surface of all specimens was mechanically ground using a grinding machine. The specimens were then ultrasonically cleaned with ethanol for 15 min and dried in a hot air drier. Before SPS, the surface of the substrate was polished using emery paper of
# 80 to #1000 and treated with grit blasting using alumina particle 30, at a spraying pressure of 5 MPa. Once the surface of the specimen was blasted to create the desired roughness, they were cleaned ultrasonically using ethanol for 20 min.

*2.4. Fabrication of Coating by SPS*

A suspension flow system was designed and assembled with a plasma gun system. Four different compositions of the feedstock were used to evaluate the effects of composition on mechanical and antibacterial properties of the SPS-HAp composite coating. The coated layer was deposited using a commercial plasma spray machine (Model 9MB, Sulzer-Metco Ltd) connected to a suspension flow system at an optimized operating condition, as shown in Table 1. The desired thickness of the suspension plasma-sprayed coating layer was 100-150 m.

*2.5. Coating characterization*

After fabrication of the coating, the morphology of the specimen was investigated using SEM (JEOL-JCM 6000 Plus, NeoScope, JEOL, Japan). Image of the surface microstructure was captured by detecting the secondary electron (High-vac) upon exciting the atoms using an electron beam at a voltage of 5 kV. The crystalline phases of the coating were identified using HXRD (Smart Lab 9KW, Rigaku, CO. Ltd. Japan) at CuK radiation, a grazing incident of $2°$, a voltage of 45 kV and a current of 200 mA. HAp (ICDD 00-009-0432), rutile (ICDD 00-001-1292), anatase (ICDD 00-001-0562), $Ti_6O_{11}$(ICDD 01-076-1266), $Ti_5O_9$ (ICDD 01-071-0627), $Ti_4O_7$ ((ICDD 01-077-1390)), and $Ti_3O_5$(00-023-0606) were used in the Rietveld analysis. XRD data of pure titanium has also taken to remove the peak coming from the substrate.

Raman spectroscopy (Lab-RAM HR-800, Horiba Jobin YVON) was conducted to identify the phase distribution of the coatings. The conditions were as follows: Laser wavelength of 532 nm; laser power of 50 mW; slit size of 100 m; hole size of 500 m; objective lens magnification of x100; observation area of (100 x 100) m; exposure time of 3 s; accumulation number of 5. Raman spectroscopy of unheated HAp, $Ti_2O_3$, HAp/titania, rutile, and anatase were also investigated as reference data. In addition, phase analysis of the coating was performed via XRD Rietveld and Raman peak fitting analysis using Gaussian equation. The contact angle was also measured using 1 L droplets.

*2.6. Interfacial strength test*

Interfacial test following ASTM F1147-05 & JSME standard S019 was conducted using the autograph (AGS-X 10N-10KN Shimadzu co. Ltd. Japan) with a cross-head displacement rate of 1 mm/min. Adhesion of the specimen using the metal lock Y610 was performed following the method proposed by Otsuka et al.[43]. A strain gauge was attached at the bottom of the specimen 10 mm from the coating layer. To detect the failure of the coated layer, AE system with fast Fourier analysis (FFT) were used with a total gain of 80 dB and $V_L$ & $V_H$ were respectively 0.4 and 1 V. Interfacial strength was calculated by dividing the maximum load by the nominal area.

Fracture surfaces (both the adhesive side and the coating side) were observed using a digital microscope and images of the fracture surface were obtained. Raman spectroscopy was conducted on the coated side to confirm the absence of infiltrations of the adhesive into HAp/Gray Titania coating during the tensile test. Infiltration was prohibited to obtain valid values for the bonding strength of the coating.

*2.7. Hardness and Young modulus test*

Hardness and young modulus were determined using dynamic microhardness test-ing machine (DUH-211S, Shimadzu, Japan). Testing was conducted on polished cross-sectional surfaces of the embedded specimen under a load of 200 mN. Time set for both loading and unloading was 15 s and holding time was 5 s. A total of ten points were selected for each type of coating to obtain the hardness and young modulus values. Mean and standard deviation of data from ten points was subsequently calculated.

*2.8. Fracture toughness test*

Fracture toughness was evaluated using the indentation fracture (IF) method with a dynamic microhardness testing machine (DUH-211S, Shimadzu, Japan). To evaluate $K_{IC}$ using the IF method, several equations were proposed, some of which required of young and Poisson modulus values in addition to hardness values. Niihara's equation, 1, was used to calculate the fracture toughness value.

$$K_{IC} = 0.067(E/Hv)^{0.4}(Hv)(alpha)^{0.5}(c/a)^{1.5} \quad (1)$$

$K_{IC}$ : Fracture Toughness (MPam$^{1/2}$)

H : Vickers Hardness [MPa]

P : Test Load in Vickers Hardness Test [MPa]

E : Young Modulus [MPa]

a : Half average length of the diagonal of Vickers marks [$m$]

c : Average length of the crack obtained in the tip of the Vickers hardness marks [$m$]

*2.9. Antibacterial test*

The antibacterial test was conducted using E. coli (K-12 strain, NBRC 3301, National Institute of Technology and Evaluation, Japan) to investigate the effect of the composition of SPS-HAp/Gray Titainia coating on antibacterial properties. CFU values were counted to estimate the number of viable bacteria in a sample. Three types of visible light, blue, green, and red, were used to irradiate the samples for 1 h, keeping the intensity value of 50 mW/cm$m^2$. The wavelength of the three types of visible light was measured using a spectrometer (MK 350N premium) before the test. The Luria Bertani (LB) medium consisted of Bacto tryptone (10 g/L), Bacto yeast extract (5 g/L), NaCl (10 g/L), and deionized water. Agar medium consisted of Bacto tryptone (10 g/L), Bacto yeast extract (5 g/L), NaCl (10 g/L), Bacto Agar (15g/L), and deion-ized water. Both media were then sterilized in an autoclave at 120$^o$ C and 15 psi for 2 h. (TOMY, SX-500) The initial concentration of the of E. coli suspension was 10$^5$ CFU/ml. Droplets of E. Coli suspension (100 l) placed on the surface of the coating (10 x 10 mm) following the international standard for antibacterial tests. Visible light was irradiated from the top of the coating for 1 h. The E. coli suspension, following the light irradiation, was diluted 10$^4$ times. After that, 0.1 mL of the diluted suspension was grown on agar medium and incubated at 37$^o$ C for 18 h, as shown in Figure 1. Finally, Colonies on the nutrient agar medium were counted using AI camera. CFU was calculated.

Cell viability was evaluated to reveal the intensity of the live/dead cells on the sur-face of each coating after light irradiation. E. coli nuclei were stained using SYTO 9 from Thermo Fisher Scientific Inc. following the manufacturer's instructions. Then, adhesion and viability of E. coli were visualized using a fluorescence microscope (BZ-8100,Keyence Co. Ltd.).

ANOVA and several comparison by Holm's method were applied to the result of the antibacterial test. The significance level was set to $p < 0:05$. All statistical analyses were performed using the R4.0.3 software.

3. Results

3.1. Characterization of feedstocks

Figure 2 (a & b) shows the typical surface morphology of the submicron-sized powders. The microscopic image depicts the homogeneous distribution of the particle with little agglomeration. Moreover, the agglomerated powders could be dispersed easily in the suspension. Figure 2 (c) shows particle size distribution of HAp and $Ti_2O_3$ powder. The mean size of HAp was 0.775 0.1 m and that of $Ti_2O_3$ was 0.535 0.4 m. This result certified that powder size was much smaller than the ones used for a conventional APS technology, which were more than 45 m. Larger variation in the sizes of $Ti_2O_3$ powders were probably due to hardness, which would resist to fractures during the ball milling process.

3.2. Coating features deposited by SPS

Figure 3 shows the surface morphologies of for all four types of coating deposited using SPS. The suspension plasma-sprayed coating exhibited white color for pure HAp coating and became more black with the increasing concentration of $Ti_2O_3$. Only in the case of HAp 50:$Ti_2O_3$ 50 showed gray color, which might be associated with higher heat consumption during the coating deposition process. Figure 4 (a-d) shows the sur-face microstructure of HAp/Gray Titania coatings deposited by SPS at optimized con-ditions. The surface of the HAp/Gray Titania coating showed homogeneous features with low porosity. Figure 4 (e-h) depicts the highly magnified SEM images of the coat-ings. Splats were flattened and few spherical splats were observed, which suggested that particles were sufficiently melted during SPS [44]. The splats were well connected, which was expected to provide high cohesion strength of the coatings. Porosity distri-bution and thickness of each type of coating are summarized in table 2. The thickness of the coating layers was 70-100 m, which was compatible with the ones fabricated using the conventional APS method. Porosity of the coatings decreased with the in-creasing concentration of $Ti_2O_3$ in the suspension.

*3.3. Phase identification of the coating*

The microstructure of HAp/Titania coating was also characterized using Raman spectroscopy and HXRD to identify the crystal structure and phase distribution of the coating. Figure 5 (a-d) shows the XRD patterns of the coating deposited by SPS. Though the primal peak at approximately 2 = 32.0 degree had a shoulder due to for-mation of -TCP, the main phase showed HAp with higher crystallinity [45]. For four types of composite coating, $Ti_2O_3$ was thermally oxidized into anatase or rutile ($TiO_2$). SPS could decompose $Ti_2O_3$ into mixed Magneli phases of $TiO_2$. $Ti_4O_7$ and $Ti_3O_5$ phases were detected, which usually shows photocatalytic activity by visible light irradiation [46]. Matsuya et. al. observed the different phase $Ti_6O_{11}$ in their composite coating fabricated by APS [25]. SPS produced the Titania coating with Magneli phases effectively, which enhanced antibacterial properties.

Figure 6 (a-d) shows the Raman spectroscopy results of coatings. Pure HAp coating had the main peak from 961 cm$^{-1}$, which was due to the $_1$ stretch of $PO^3_4$ in the HAp crystal. Besides the peak from HAp, we also observed peaks from rutile and anatase phases of $TiO_2$, which were compatible the ones observed using unheated rutile and anatase $TiO_2$ powders. Raman spectroscopy results also indicated the existence of Magneli phases by showing a strong peak of $Ti_4O_7$ and $Ti_3O_5$. The vibrational frequencies at 240, 247, and 605 cm$^{-1}$ were from $Ti_4O_7$ and the frequency at 265, 615 cm$^{-1}$ were from $Ti_3O_5$. Consequently, SPS produced composite coating, contain-ing oxygen-deficient types of titania, which could be used as a visible light responsive photocatalyst. In this study, Gaussian fitting was used to study the detailed bonding structure of titanium and oxygen. Peak distribution from 500 to 800 cm$^{-1}$ was fitted to analyze the available phases. However, the peak produced by $PO^4$ due to the HAp crystal was the obstacle for detecting the peak solely obtained from the phases of $TiO_2$. Hence, the substitution of the peak from HAp (both primary and secondary peaks) was performed to avoid all the peaks from the HAp coating.

Volumetric composition of Magneli phases for all three types of coating was also evaluated using both Raman multiple peak fitting analysis and XRD Rietveld quantitative analysis (Figure7). Both Raman and XRD results consistently exhibited that Magneli phases were increased with the increasing nominal composition of $Ti_2O_3$.

*3.4. Interfacial strength via tensile testing*

Figure 8 (a) shows the stress-strain curve during interfacial strength tests. The stress-strain curve showed a linear line. AE signals were detected only in the final stage close to fracture. The number of detected AE signals were reduced compared to the case of HAp coating deposited by APS, which was due to reduced porosity. Interfacial strength of HAp coating was approximately 15 MPa, which was higher than that of HAp coating by APS(approximately 10 MPa) [43] and the strength for composite coating was slightly increased with the increasing $Ti_2O_3$ concentration (Figure 8 (d)). There was a little time lag in detecting two AE signals during the tests, which suggested that the damage occurred at the midpoint of two sensors; in the coating or ad-hesive layer of both HAp and composite coatings (Figure 8 (b)). FFT analyses showed that the main frequency domain was approximately 500 kHz for both HAp and com-posite coatings (Figure 8 (c)). Such AE signals were usually observed in the cracking of HAp coating, which suggested the validity of interfacial tests [47]. Infiltration of adhesive was abstained, which certified that the strength detected using the tensile test was the interfacial strength of the coating with substrate.

*3.5. Hardness and Young modulus test*

Figure 9 (a) and figure 9 (b) show the effect of $Ti_2O_3$ addition in suspension on mechanical properties of HAp/ Titania composite coating. Both values were significantly increased with the increasing concentration of $Ti_2O_3$ in the suspension due to homogeneous dispersion of harder $TiO_2$ particles in the coatings.

### 3.6. Fracture toughness test

The result of the fracture toughness test using IF method is depicted in figure 9 (c). Pure HAp coating exhibited a fracture toughness of 1.14 MPa$m^{1/2}$. Addition of $Ti_2O_3$ enhanced the fracture toughness probably due to reduced porosity and fewer number of microcracks presented in the composite coating.

### 3.7. Antibacterial activity of HAp/gray Titania coating

Results of the wavelength for three types of light is shown in figure 10. It was observed that wavelengths of the blue, green, and red lights were 447, 518, and 635 nm respectively. CFU for various compositions of HAp/Gray Titania coating was evaluated as depicted in figure 11. CFU of the SPS 80H20T sample was approximately 20 % lower than that of the pure HAp sample. There was no reduction in CFU with no sample, which suggested that the enhanced antibacterial property was not attributed to heating by irradiation but to photocatalytic effects of the SPS-HAp/Gray Titania coating. The ruducion effects on CFUs by titania containing groups (80H20T,60H40T,and 50H50T) and laser irradiation were significant ( $F_{(4)}$ = 72.95, $p < 0.001$, $F_{(3)}$ = 16.4, $p < 0.001$) ,whereas an interaction effect (*group*) (*laser*) was weakly significant ( $F_{(12)}$ = 1.85, $p = 0.072$). Detailed analyses by multiple regression analyses revealed that the inteaction effects of titania containing groups(60H40Tand 50H50T) with light irradiation (Blue,Green, and Red) were significant ( $t$ = 2.98, $p = 0.005$( 50H50T Red irradiation)). These results certified that the new SPS composite coating could success-fully improve the antibacterial properties. CFU was further decreased by increasing the weight percentage of the antibacterial agent (Gray titania:$TiO_2$ containing magneli phases).

The antibacterial efficiency of the SPS composite coating was also assessed via the live/dead bacterial viability assay. As red light irradiation showed remarkable reduc-tion in CFU, we checked cell viability only in the presence of red light irradiation for all four types of coating. As shown in figure 12 (b-e), the number of live bacteria (green) on pure HAp coating was more than that on other composite coatings, and almost no dead cells (red) were found. The intensity of the live/dead cell ratio for all four types of coating was calculated as shown in figure 12 (a). This result also confirmed that the bacterial growth was inhibited on the surface of the composite coating. Consequently, the addition of $TiO_2$ enhanced the antibacterial property of the SPS composite coating under visible light irradiation, and 50H50T coating showed the maximum efficiency in this regard.

*3.8. Contact angle of coating*

The contact angle was measured for all four types of coating, and pure titanium. Figure 13 shows that HAp coating exhibited hydrophilic behavior, whereas the composite coating was not hydrophobic. Meaning that SPS composite coating resulted in hydrophilic surfaces. However, wettability was decreased with higher concentration of $TiO_2$.

4. Discussion

This study aimed at observing the effects of composition on mechanical, and antibacterial properties of HAp/ Gray Titania coating fabricated by SPS. New suspension-feeding system was successfully attached to the conventional APS gun, which enabled us to directly compare the mechanical properties of plasma-sprayed coating fabricated using different feeding methods, air or suspension. SPS could produce a homoge-neous composite coating with lower porosity, which enhanced hardness, young's mod-ulus, and fracture toughness by increasing the concentration of $Ti_2O_3$ in the suspen-sion(Figure 4, Figure 8). Interfacial strength of the composite coating was slightly increased with the increasing concentration of $Ti_2O_3$ in the suspension. (Figure 9).SPS specifically suppressed the oxidation of $Ti_2O_3$, and the resultant composition of the coatings successfully contained Magneli phases of titania (Figure 5,Figure 6, Figure 7), which could enhance photocatalytic property of the composite coating under irra-diation with visible light(Figure 11,Figure 12). Probable enhancement mechanisms in mechanical and antibacterial properties of SPS-HAp/Gray Titania coating are discussed below.

*4.1. Enhancement of mechanical properties of the SPS-HAp composite coating* Tensile test results demonstrated that the strength of the HAp coating was approx-imately 15 MPa, which was significantly higher than the that of the HAp coating deposited using conventional APS method [43]. The higher strength of the SPS-HAp coating might be attributed to lower particle size, lower porosity, and higher crystallinity. Fine particles were partially melted, which improved the bonding strength among splats and reduced porosity. $TiO_2$ dispersing was possibly due to unchanged bonding conditions between both layers. The reinforcement of $Ti_2O_3$ in HAp significantly enhanced the hardness, young modulus and fracture toughness of the coating. The cohesive strength of HAp coating was significantly improved by dispersing $TiO_2$. Higher crystallinity of the HAp coating, which was attributed to evaporation of the solvent during SPS, could increase cohesive strength of the coating layer because of suppressed formation of the amorphous interface layer. Therefore, the HAp coating deposited by the novel SPS method showed higher interfacial strength than APS coat-ing.

*4.2. Enhancement mechanism of antibacterial properties of the SPSed HAp composite coating*

In conventional APS process, HAp particles were impacted on the substrate and then cooled down, which led to the formation of thermally decomposed phases or amorphous phases, such as tri calcium phosphate, tetra calcium phosphate, and CaO, which reduced crystallinity of the coating layer. [48]. Such lower crystallinity and low coating strength were not beneficial for the mechanical integrity *in vivo*. in contrast, during SPS process, the temperature gap between molten material and the substrate was lower due to heat consumption by the solvent. Such exploitation of heat by evap-oration of the solvent helped reduce the cooling rate of the particle and subsequently reduced the formation of the amorphous phase, which led to higher crystallinity of HAp coating [49]. In addition, composite coating contained several Magneli phases, such as $Ti_4O_7$ / $Ti_3O_5$, which showed stronger photocatalytic activity than other Magneli phases detected in previous studies [25]. Our present study revealed that SPS could provide $TiO_2$ coating with Magneli phases. The band gap energy of Magneli phases is approximately 0.5 eV narrower than that of anatase $TiO_2$ (3.2 eV) [50]. Morakul et. al. have demonstrated that the composite coating of HAp with titania exhibits antibacterial properties against *E. Coli* under visible light irradiation, such as blue(450nm=2.76eV), green(532nm=2.33 eV), and red (633nm=1.96eV) [21]. Formation of such Magneli phases could contribute the enhanced antibacterial properties using visible light.

Enhancement mechanism using fine particles in HAp composite coating is illus-trated in figure 14. In the case of SPS composite coating, the grain size was around 2 m, which could only accommodate 1 or 2 *E. coli* bacteria, which increased the contact distance between coating particle and bacteria, so that the bacteria was more exposed to released reactive oxygen species (ROS). On the contrary, for conventional composite coating, the grain size was quite big (10 m ), and the number of adhered bacteria on each grain was more than that obtained using SPS coating. It is not practically possible for each radical to go through a large number of bacteria. As a result, many bacteria did not come in contact with ROS and remained alive. Contact angle of the SPS-HAp coating was superhydrophillic, which also played a supportive role to enhance bacte-rial adhesion onto coating surfaces. Consequently, the particle size in the coating was a considerable factor in enhancing the antibacterial property of HAp composite coating.

In this study, SPS could successfully modify the microstructure and crystallo-graphic structure of the coating material by utilizing the nanoparticle in the form of suspension, which was favorable for antibacterial properties. The solid concentration (weight percentage of the particle) and solvent (both aqueous and alcoholic) played an important role on the microstructure of the coating. Optimizing such parameters of SPS led to the control of the coating microstructure. On the contrary, cortical bone on the surface provides higher mechanical strength due to their complex dense structure, and cancellous bone inside the enables blood circulation and bone ingrowth attributing to their porous structure[51]. Both composition and porosity gradient layers should be combined in single coating structure in order to mimic the structure of human bone. Outer layer should be porous, made up of HAp, whereas the inner layer should be compact containing the metal. The present study does not fulfill the requirement of the ideal coating to be applied in implants. Therefore, it is further important to challenge functionally graded coating for biomedical application[52].

5. Conclusion

This study investigated the effects of composition on mechanical and antibacterial properties of hydroxyapatite /gray titania coating fabricated by suspension plasma spray. The main conclusions obtained are as follows:

1. SPS successfully deposited composite coating of HAp with $Ti_2O_3$ of different compositions onto the surface of the titanium substrate. The microstructure of the coating certified that particle distribution was homogeneous, most of the splats were flatten and melted properly, which resulted in low porosity.
2. HAp coating showed very high crystallinity and composite coating contained mainly rutile and anatase phases. SPS also provided $TiO_2$ coating with oxygen-deficient types of titanium oxides. *e.g.*, Magneli phase, which can be activated by visible light irradiation.
3. The enhanced antibacterial property of the SPS-HAp/Gray Titania coating can be attributed to high composition of Magneli phases and high wettability.

This study revealed that finer particles in the microstructure of the HAp/ Gray Titania promote visible light sensitivity, which provides superior antibacterial properties. Optimization of process parameters and coating layers should be further considered in future studies.

Acknowledgments

We thank Niigata Metallicon Co., Ltd., for producing the plasma-sprayed samples. We also thank the analytical center of Nagaoka University of Technology for their supports in XRD and SEM observations. This study was partically supported by JSPS KAKENHI 17H04898 ,Union Tool Foundations and TAKEUCHI foundations.

List of Figures

List of Tables

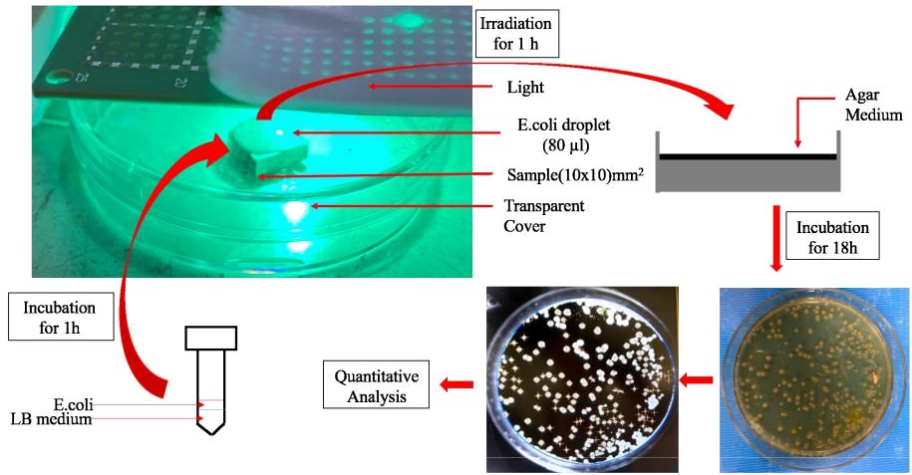

Figure 1: Experimental Procedure of Antibacterial Test

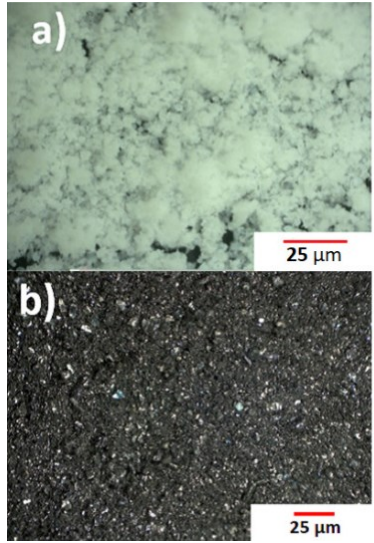
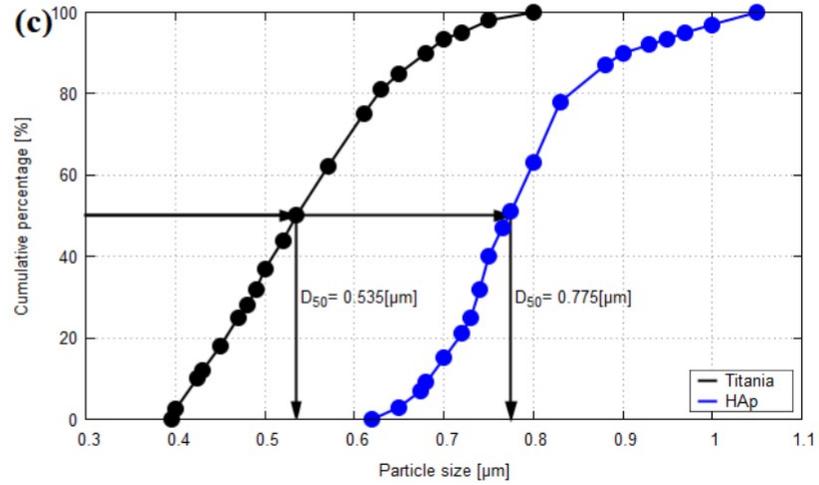

Figure 2: Microscopic Image of the Submicron Sized Powder Used as Feedstock in SPS process (a) Hydroxyapatite powder (b) $Ti_2O_3$ Powder (c)Cumulative Size Distribution of the HAp Powder(blue) and $Ti_2O_3$ Powder(black) for Suspension Plasma Spray

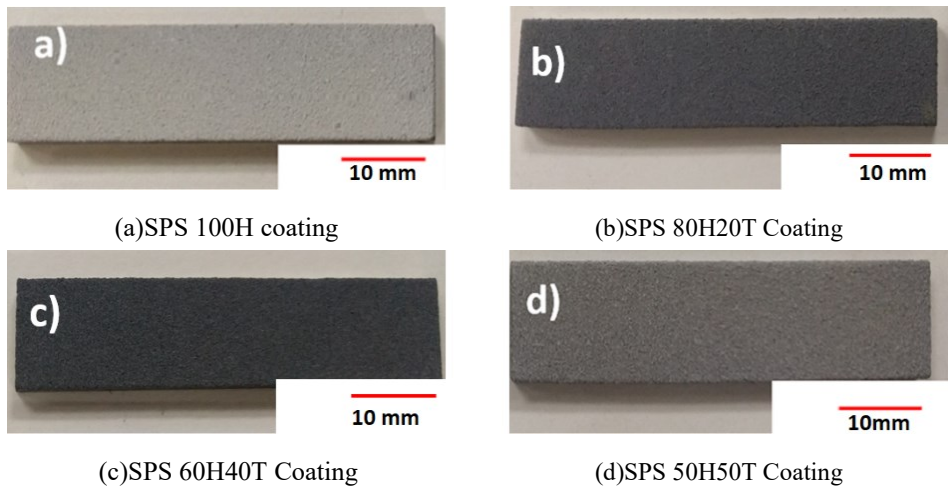

(a)SPS 100H coating

(b)SPS 80H20T Coating

(c)SPS 60H40T Coating

(d)SPS 50H50T Coating

Figure 3: Surface Features of SPS Coating on the Surface of Commercially Pure Titanium Material

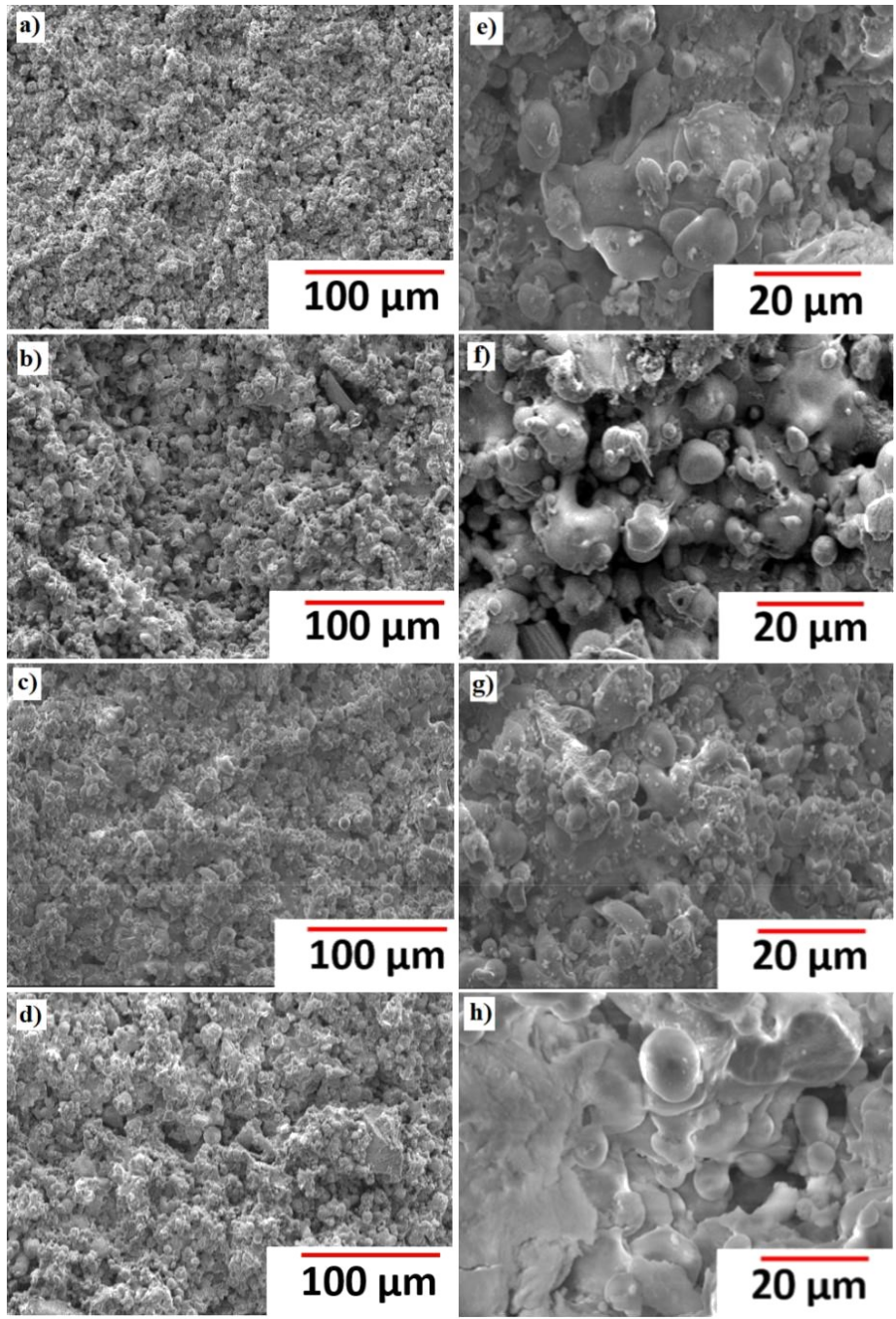

Figure 4: Surface Morphology of the Coating Deposited by SPS technology at an optimized condition; (a), (b), (c) & (d) are the coating surface of 100H, 80H20T, 60H40T & 50H50T coating respectively. (e), (f), (g) & (h) are the corresponding higher magnification image.

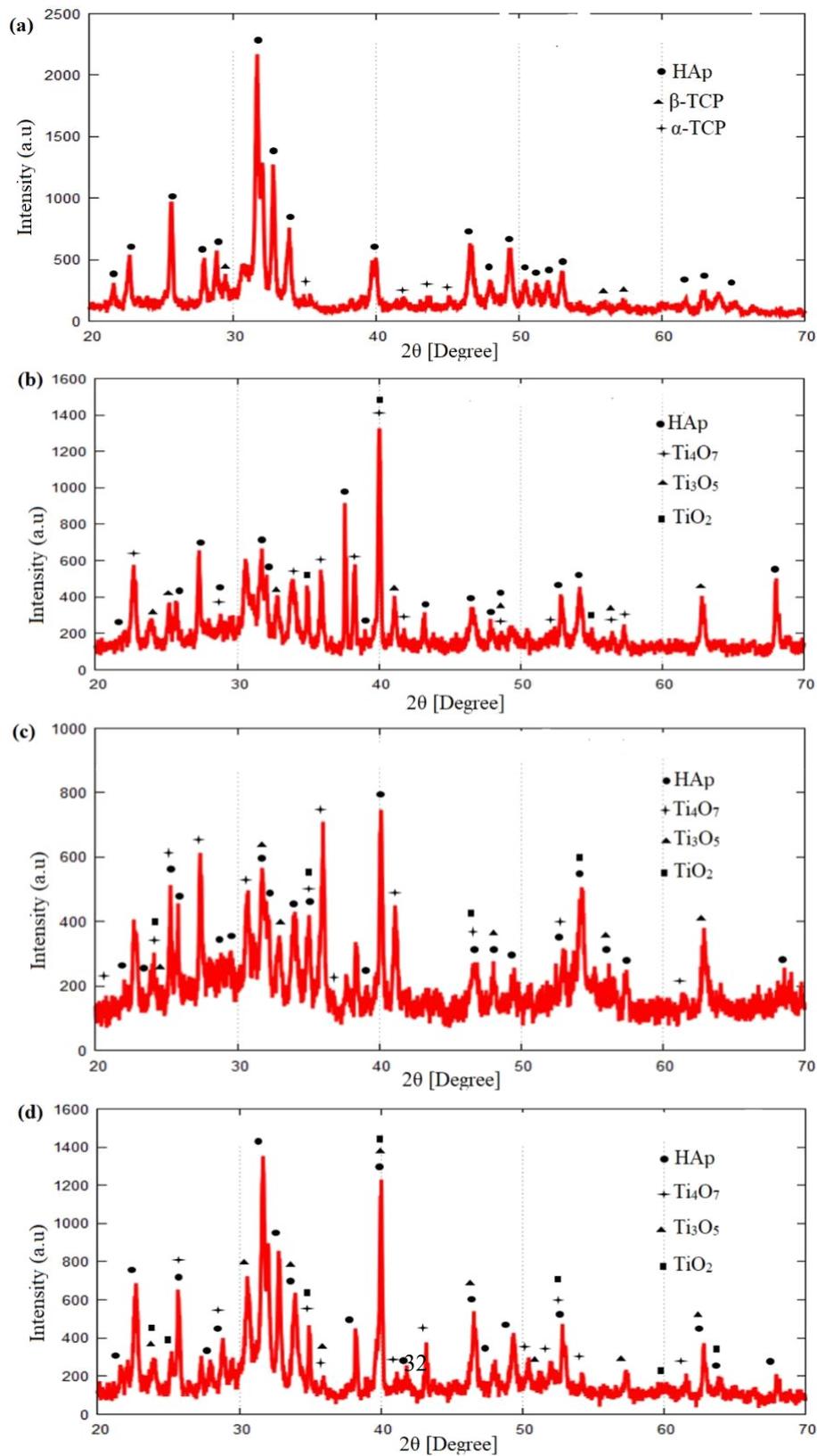

Figure 5: Phase Identification of the Various Composition SPS Coating by XRD (a) 100H (b) 80H20T (c)

60H40T (d) 50H50T

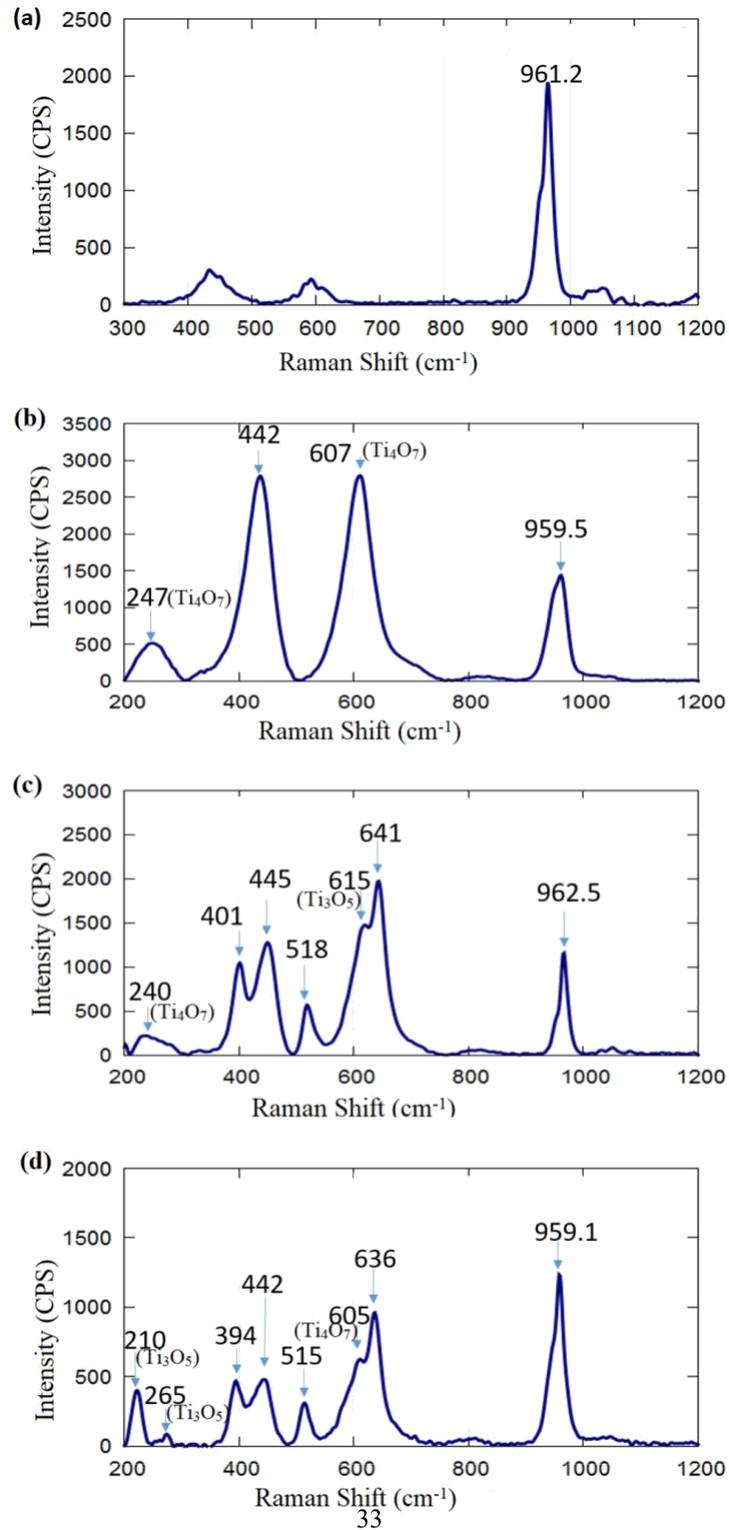

Figure 6: Phase Identification of the Various Composition SPS Coating by Raman Spectroscopy (a) 100H (b) 80H20T (c) 60H40T (d) 50H50T

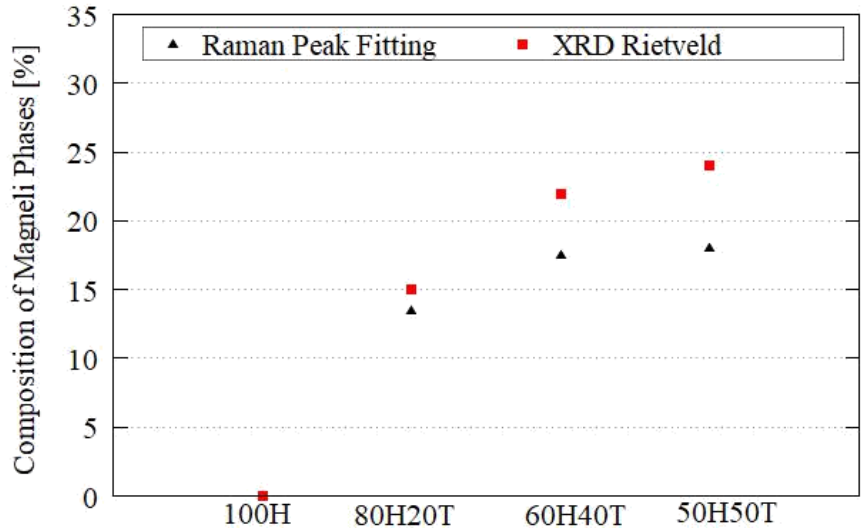

Figure 7: Quantitative Evaluation of Magneli Phase by Raman Peak fitting analysis and XRD Rietveld analysis Contained in Each Type of SPS coating

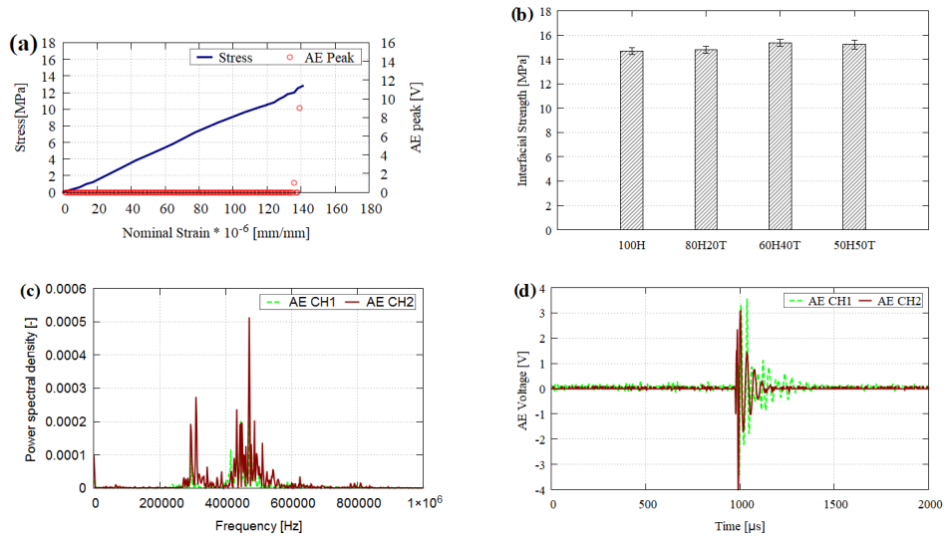

Figure 8: (a)Stress-Strain Curve and Corresponding AE Signal during Interfacial Strength Test for SPS coat-ing. Detected AE SIgnal During Interfacial Srength Test (b) Typical AE Signal at Fracture (c) Corresponding FFT Signal. (d) Effect of composition on Interfacial Strength

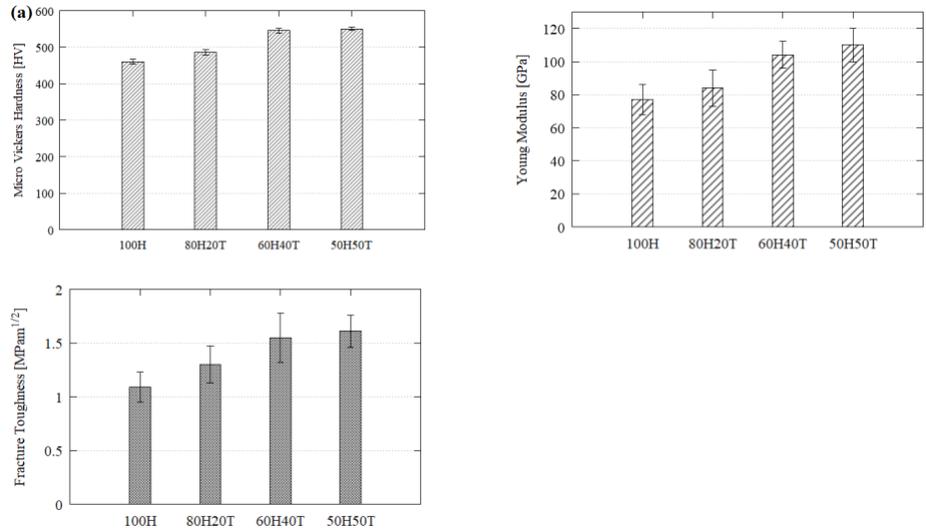

Figure 9: Effect of Composition on Mechanical Properties of Suspension Plasma Sprayed HAp/ Titania Coating (a) Hardness (b) Young Modulus (c) Fracture Toughness

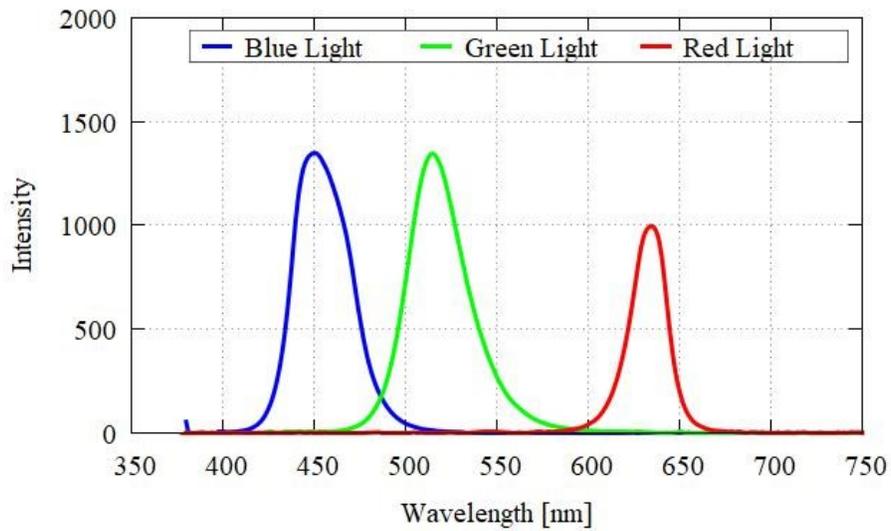

Figure 10: Wavelength Distribution Measured by Spectrometer of the Blue, Green, and Green Light; Used for Antibacterial Test

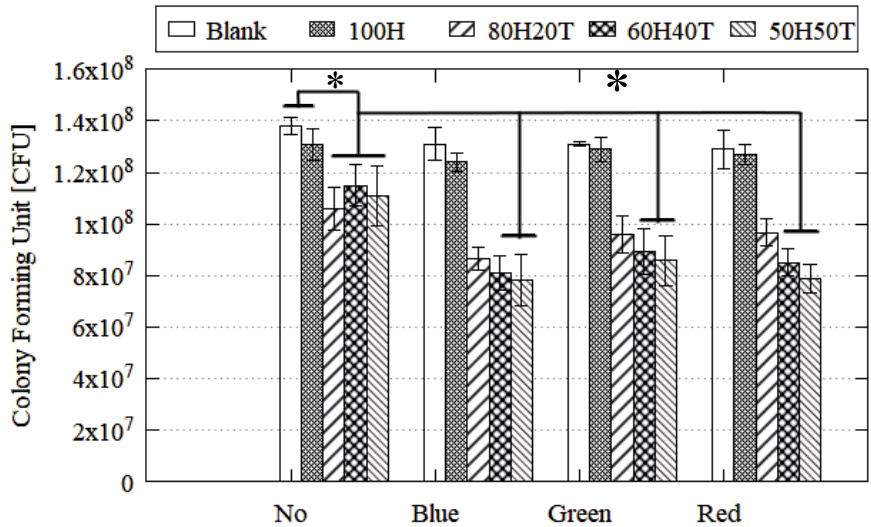

Figure 11: Effect of Composition on Colony Forming Unit (CFU) Values of Suspension Plasma Sprayed HAp/ Titania Coating. Interaction effect of composition and laser irradiation was significant by two-way factorial ANOVA method. (Star symbol indicates that the value of p is less than 0.05)

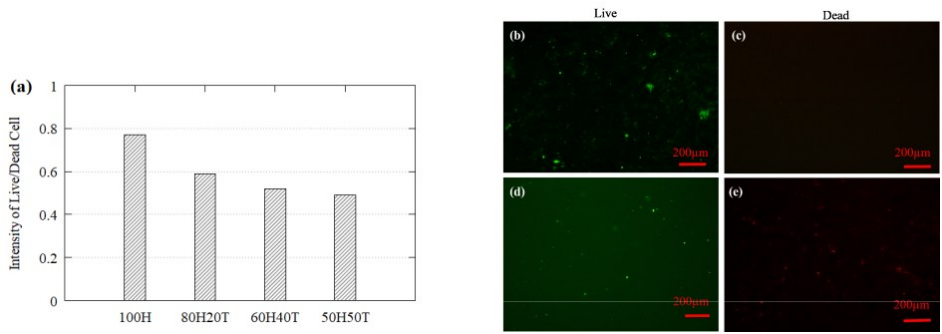

Figure 12: (a) Quantitative histogram analysis of the density of live bacteria; Live/dead staining of E. coli cell on the surface of SPS coating (b) 100H (live cell) (c) 100H (dead cell) (d) 50H50T (live cell) (c) 50H50T (dead cell)

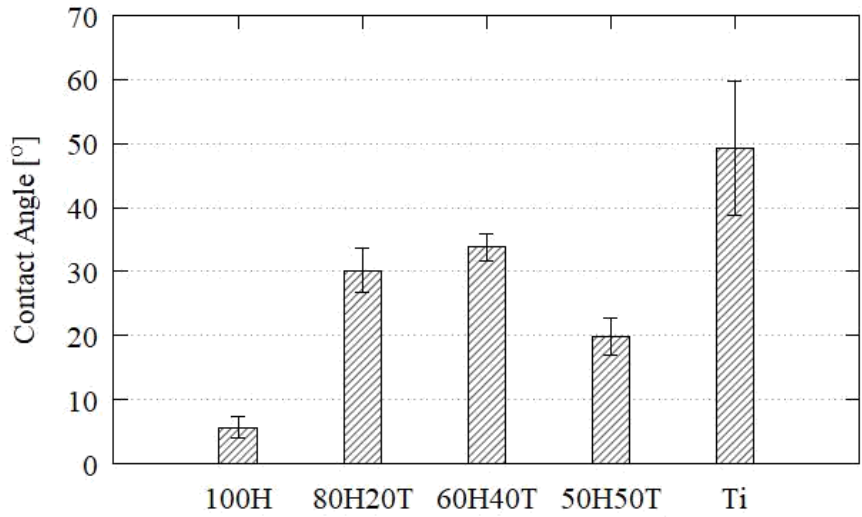

Figure 13: Wetability of the SPS coating was observed by measuring the coantact angle of each type of coating

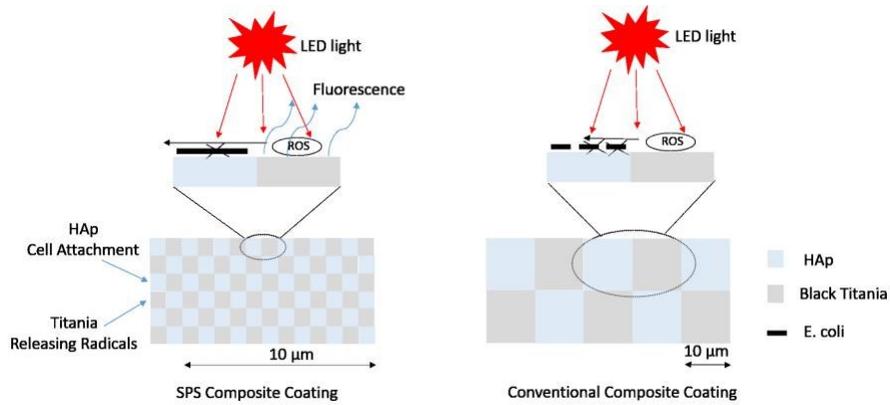

Figure 14: Comparing bacterial inhibition process during light irradiation between SPS and conventional composite coating

Table 1: Optimized Operating condition of SPS method for fabrication of HAp composite coating

| Suspension Flow System | | Plasma Spray | |
|---|---|---|---|
| Parameter | Value | Parameter | Value |
| Solvent Type | Water + Ethanol | Current(A) | 600 |
| Injection Pressure(MPa) | 0.5 | Voltage(V) | 60 |
| Nozzle ID(mm) | 0.52 | Spraying Distance (mm) | 120 |
| Flow Rate (ml/min) | 55 | Shield Gas | Ar |
| Solid Concentration (wt. %) | 20,25 | Coating Thickness ($\mu m$) | 100-150 |

Table 2: Thickness and Porosity Distribution of the Coating Deposited by SPS

| Coating Type | Thickness ($\mu m$) | Porosity (%) |
|---|---|---|
| SPS 100H | 75 ± 10 | 11.4 ± 2.9 |
| SPS 80H20T | 90 ± 22 | 10.3 ± 1.9 |
| SPS 60H40T | 105 ± 25 | 7.5 ± 2.1 |
| SPS 50H50T | 100 ± 20 | 8.1 ± 3.1 |